# Observation of double superconducting transition in $Y_{0.55}Pr_{0.45}Ba_2Cu_3O_{7-\delta}$ polycrystalline sample by Hall effect measurement

*M. S. da Luz, C. A. M. dos Santos\*, and A. J. S. Machado*

*Grupo de Supercondutividade, Departamento de Engenharia de Materiais,*

*FAENQUIL, 12.600-970, Lorena-SP, Brazil*

## ABSTRACT

This work reports longitudinal ($R_{XX}$) and transverse ($R_{XY}$) resistance as a function of the temperature measured at low magnetic fields for a $Y_{0.55}Pr_{0.45}Ba_2Cu_3O_{7-\delta}$ polycrystalline sample. It is observed nonzero transverse voltage at zero external magnetic below superconducting transition and above no such voltage was detected. Comparing $R_{XX}$ (T) and $R_{XY}$ (T) curves it was possible to observe a correlation between Hall resistance and double resistive superconducting transition. Magneto-resistance $R_{XX}$(H) measurements showed that the dissipation is dominated by weak coupling between superconducting clusters. I-V and R(T) curves of both components suggest that the positive Hall voltage at superconducting state must be related to quasiparticle currents.

\* Corresponding author. Fax: +55-12-553-3006

*E-mail address*: cams@demar.faenquil.br

## 1 - INTRODUCTION

During the last decade great attention has been devoted to the theoretical and experimental studying of the Hall resistance in the superconducting state. The most interesting behavior observed in many high critical temperature ($T_C$) and in some conventional superconductors is the sign reversal in the Hall voltage which crosses from positive at normal state to negative near the critical temperature [1-13]. According to the classical model, the Hall effect in the mixed state should have the same sign as that in the normal state due to a normal current flowing through the vortex core [14-16]. There are several models to explain the sign changing of the Hall resistance [3-5,7,8], but no agreement has been achieved [17]. Some questions remain in discussion, if Hall effect is an intrinsic electronic property or it is related to vortex effect such as individual or collective motion and if pinning effects are important. Experiments have revealed that this Hall anomaly has been observed in moderate magnetic fields and must be connected with carrier concentration [17].

Another important feature of the Hall effect in the superconducting state is the scaling of the transverse and longitudinal resistivity, $\rho_{XY} \sim (\rho_{XX})^\beta$, which has been observed occur with $\beta \sim 1.7$ [17,18]. An universal scaling law was derived near a vortex-glass transition with the specific prediction of $\beta = 2$ [19-21]. This scaling model predicts that the Hall conductivity should be independent of pinning effects, therefore, there are several reports regarding samples with artificially introduced defects which discuss the pinning dependence of $\sigma_{XY}$ [17,21-23].

Additionally, a second sign reversal of the Hall resistance has been observed in several samples of high critical temperature superconductors [17,24-28]. In this double sign reversal, the sign of the Hall resistivity is positive in the normal state, changes to negative near $T_C$ and becomes positive again at lower temperatures. The origin of this effect remains in great discussion [17,27,28]. In order to explain this phenomenon, Göb *et al.*[17] have considered the Hall conductivity ($\sigma_{XY}$) into three components related to the quasiparticle or vortex-core contributions which is associated with the normal state excitations ($\sigma^N_{XY}$), superconducting dissipation resulting from hydrodynamic vortex effects and superconducting fluctuations ($\sigma^S_{XY}$), and pinning dependence ($\sigma^P_{XY}$). The Hall conductivity is calculated using these three components in the following way:

$$\sigma_{XY} = \sigma^N_{XY} + \sigma^S_{XY} + \sigma^P_{XY} \ . \qquad (1)$$

More recently great attention has been given to the appearing of a transverse voltage at zero external magnetic field in HTSC near superconducting transition [29-34]. Glazman [31] proposed a model for the explanation of this effect in which the magnetic field produced by the current crossing the sample can penetrate into it in the form of vortex of different sign. This sign is determined by different direction of magnetic field on the opposite edges of the sample resulting in vortex penetration from the opposite one. These vortices are called vortex and antivortex which move in opposite direction under the influence of Lorentz force and can annihilate each other if the attractive interaction between them overcomes the Lorentz force. This means that the path of vortex and antivortex will be distorted and transverse voltage appears according to the Josephson relation. According to this theory the transverse voltage value should increase for the low transport current and again decrease in the higher current limit [31].

Vašek *et al.* [31,32] have also observed transverse voltage at zero applied magnetic field in superconducting samples. During the last years Vašek´s group introduced the idea of guiding force to explain their results. It considers the influence of the guiding force on the moving of the vortex and antivortex. Vašek *et al.* supposed that an intrinsic pinning force related to linear channels inside the sample can induce vortex and antivortex motion in a specific direction giving a nonzero macroscopic transverse voltage [32, 33].

On the other hand, effects of weak coupling between grains or superconducting clusters have been extensively studied in polycrystalline samples [35-42]. The main observed effect is the broadening of the superconducting transition which has been separated in two distinct regimes. Below the onset of the critical temperature, $T_{Ci}$, the reduction of the resistance has been related to the formation of isolated superconducting clusters inside sample. The decreasing of the temperature causes the beginning of connection of the superconducting clusters, mediated by Josephson coupling, at a temperature labeled $T_{Cj}$. This regime is non-ohmic and a further decreasing of the temperature produces an infinite network of connected superconducting clusters which leads the sample to a true zero-resistance superconducting state. This effect has been often related to a branching point in curves of the electrical resistance as a function of the temperature measured under different applied currents [35-38]. The weak coupling mediated by Josephson effect between superconducting clusters has been unambiguously determined by clockwise hysteresis loops in magneto-resistance measurements [38-40].

Particularly in our group we have measured several magneto-transport properties in polycrystalline samples of the $Y_{1-x}Pr_xBa_2Cu_3O_{7-\delta}$ (Y123+Pr) and $Bi_2Sr_2Ca_{1-x}Pr_xCu_2O_{8+\delta}$ (Bi2212+Pr) [34-36,43] systems at low magnetic fields in which dissipation is clearly dominated by weak coupling mechanisms.

Based on measurements at high current limit we have suggested that the non-ohmic regime at temperature lower than $T_{CJ}$ can be described by a mechanism due to flowing of normal current in parallel with supercurrent which is in agreement with the resistively shunted junction model (RSJM) [41-44].

In order to obtain more information about the effects of weak coupling in superconducting polycrystalline samples, we have devoted efforts to measure Hall resistance, $R_{XY}$, and electrical resistance, $R_{XX}$, at low magnetic field limit.

## 2 - EXPERIMENTAL PROCEDURE

Polycrystalline samples of $Y_{0.55}Pr_{0.45}Ba_2Cu_3O_{7-\delta}$ composition were prepared by solid state reaction technique, using $Y_2O_3$, $BaCO_3$, $Pr_6O_{11}$ and CuO powders of high purity. Essentially, the powders were mixed in appropriated amounts, calcined at $800^O$C in air, compacted into pellets and sintered at $900^O$C for 48 hours. Crystalline phase was identified by x-ray powder diffractometry and in order to observe the granular structure of the sample we performed optical and scanning electron microscopy. The results revealed that the samples are single phase.

For the transport properties measurements, one sample was cut in the form of square with 7.2 mm in width and 0.7 mm in thickness as shown schematically in the inset of the Fig. 1(b). The electrical terminals were prepared using low-resistance sputtered Au contacts (~0.1 $\Omega$). The magnetic field, produced by a small copper solenoid, was applied perpendicular to the square surface of the sample. Constant current was applied by programmable current source (Keithley 220) and the corresponding voltage was measured using a nanovoltmeter (Keithley 181). The magneto-transport properties were measured by means of the van der Pauw technique [45,46] with permutation of the voltage and current contacts.

In this method longitudinal and transverse voltages can be detected by permutation of voltage and current contacts. Any pair of contacts can be used as current contacts and the remaining pair as potential ones. The Hall and longitudinal voltages can be formulated as, respectively

$$V_{XY} = [V_{AC,BD}(H) - V_{BD,AC}(H)]/2 \qquad (2)$$

$$V_{XX} = [V_{AB,CD}(H) - V_{CD,AB}(H)]/2 \qquad (2).$$

Noise thermopower effects has been eliminated by reversing the transport current (I) in the sample under study. Thus, to obtain longitudinal ($V_{XX}$) and transverse voltage ($V_{XY}$), eight voltage signals were measured. Now the $V_{XX}$ and $V_{XY}$ components are calculated from a combination of eight records, which can be symbolically expressed as

$$V_{XX} = \{ \ [ \ V_{AB,CD}(+I) - V_{AB,CD}(-I) \ ] + [ \ V_{BC,AD}(+I) - V_{BC,AD}(-I) \ ] \ \}/4 \qquad (4)$$

and

$$V_{XY} = \{ \ [ \ V_{AC,BD}(+I) - V_{AC,BD}(-I) \ ] - [ \ V_{BD,AC}(+I) - V_{BD,AC}(-I) \ ] \ \}/4, \qquad (5)$$

It was crucial to perform the measurements of the four components in Eq. (4) and (5) in constant experimental circumstances, especially the temperature, in order to achieve reproducible results.

Data acquisition and calculation of $V_{XX}$ and $V_{XY}$ using Eqs. (4) and (5) were done in real time on a computer. Both longitudinal and perpendicular resistance components were defined by calculating V/I for each direction.

## 3 - RESULTS AND DISCUSSION

In Figure 1 (a) is shown the electrical resistance as a function of the temperature measured at different applied currents for zero applied magnetic field for the $Y_{0.55}Pr_{0.45}Ba_2Cu_3O_{7-\delta}$ polycrystalline sample. One can see that the onset of the critical temperature ($T_{Ci}$) is close to 42.5 K and the width of the transition is as large as in other highly Pr doped Y123 polycrystalline samples [41,42]. Two distinct behaviors can be seen below the critical temperature $T_{Ci}$. The first regime, between $T_{Ci}$ and $T_{Cj}$, is ohmic and can be related to the formation of isolated superconducting clusters which reduce the equivalent resistance of the sample with decreasing the temperature. Below $T_{Cj}$ the resistance curve broads gradually with increasing the applied current. This current dependence has been suggested to be related to the Josephson coupling between superconducting clusters which leads the samples to a true superconducting state at low applied current with decreasing temperature [35-40, 42]. We have also confirmed these $T_{Ci}$ and $T_{Cj}$ values by calculating dR/dT for the both components. In the Fig. 1 (b) the results of the longitudinal resistance as a function of temperature measured under different applied magnetic fields from zero up to 48 Oe are shown. In this figure one can see the broadening of the superconducting transition increasing applied magnetic field.

To confirm that the dissipation is indeed related to the coupling between superconducting clusters we performed magneto-resistance measurements at low temperatures. Figure 2 shows clockwise hysteresis loops measured at 4.2 K for three different applied currents. The results demonstrate unambiguously that the dissipation at low temperature ($T<T_{Cj}$) is effectively due to weak coupling between superconducting clusters such as pointed out by several authors [38-40,42].

The Fig. 3 displays $R_{XX}$ and $R_{XY}$ as a function of the temperature for the sample with $Y_{0.55}Pr_{0.45}Ba_2Cu_3O_{7-\delta}$ composition measured under different magnetic fields (H = zero and $H_A$ = 4.8 Oe). From Fig. 3 (a) one can clearly see that below critical temperature nonzero transverse voltage appears at zero applied magnetic field. We can observe a sign change (to negative) below Tci. Moreover, a sign change from negative to positive with decreasing temperature is also observed (see arrows). The fact of observing Hall resistance in zero applied magnetic field is shocking. In the mixed state of superconductors, the Hall resistance is resulting from vortex hydrodynamics produced by the current going through sample [1-10]. In principle, at zero applied magnetic field there is no vortices and transverse voltage should be zero. However, transverse voltage in zero magnetic fields has been recently observed in different HTSC materials [32,33]. Near $T_C$, it is observed the existence of free vortex in the mixed state which can be generated without application of external magnetic field. In superconductors, vortex-antivortex pairs may be excited as a result of the thermal fluctuation, or an induced by transport current thought sample on the opposite sides of it. Our results can be understood based on existence of induced vortex-antivortex pairs and on application of the guiding vortex model [33]. The theory of the guiding of the vortices was proposed to explain the even effect in superconductors which supposes the existence of new force acting on the vortex. This force, called guiding force, impels the vortex to move only in a given direction that is determined by the direction of the pinning potential valley. However, the nature of this pinning potential form has not been completely understood. One of the promising models is the intrinsic pinning model which supposes that the origin of guiding forces is due to layered structure of the HTSC [32]. Other mechanisms, as grain boundary guiding in polycrystalline materials for example, should be kept in mind. This model appear seems to offer a consistent explanation of our results. The results for $H_A = 4.8$ Oe are presented in Fig. 3 (b). In the normal state (T>$T_{Ci}$) the transverse resistance presents positive signal and in the superconducting state can be seen the double sign reversal which is in agreement with other results for high-$T_C$ superconductors.

Another important observation in longitudinal and transverse resistance measurements is the fact that at $T_{Ci}$ and $T_{Cj}$ the Hall resistance changes sign (see Fig. 4). This fact is interesting and was also observed for other superconducting materials [34]. The Fig. 4 shows $R_{XX}$ (upper) and $R_{XY}$ (lower) curves measured at different applied currents under applied magnetic field of 4.8 Oe. The non-ohmic behavior in the $R_{XX}$ component at temperatures lower than $T_{Cj}$ is correlated to a positive sign of the Hall resistance which also exhibit current dependent behavior. The crossover from positive to negative in the Hall voltage at the superconducting state increasing temperature was observed to occur essentially at the same temperature of the branching point ($T_{Cj}$) in the $R_{XX}$ component (see arrows in inset of the Fig. 4). Between $T_{Ci}$ and $T_{Cj}$, the $R_{XX}$ component is not current dependent which indicates that the sample reached an ohmic state in which superconductivity is localized inside superconducting clusters [42]. At this temperature range, we also observed that the $R_{XY}$ component is negative and it does not also exhibit current dependence. The origin of this behavior in the transversal component is not understood yet but represents an important aspect related to the intragranular superconducting transition. Finally, at $T = T_{Ci}$, the $R_{XY}$ component goes back to the positive sign as expected for the hole-doped high critical temperature superconductors [1-10].

Based on results concerning the dissipation in granular superconductors described within RSJM [41,44], the $V_{XX}(H)$ results show that the dissipation is related to the weak coupling suggesting that the positive Hall voltage at low current limit must take into account the contribution due to the motion of single particles which can flow in parallel with a supercurrent [41-44]. In order to confirm this assumption, we present in the Fig. 5 some I-V characteristic curves for the longitudinal and transverse components measured at different magnetic fields. At high current limit (see inset) the longitudinal component presents similar shape of the I-V curves reported before [41,44]. At 4.2 K and low current limit it was observed that the positive Hall voltage increases with increasing magnetic field indicating correlations with normal particle currents. On the other hand, at high applied current a further increasing of the applied magnetic field causes a reduction of the Hall voltage which suggests the existence of competition between dissipation produced by intragranular flux motion and quasiparticle currents [42].

Now, let us to discuss the negative signal of the $R_{XY}$ component related to the intragranular transition. Between $T_{Cj}$ and $T_{Ci}$ the current flows essentially as normal current which agrees with the ohmic regime discussed in the introduction. Thus, $R_{XY}$ component should present only effects of the normal carriers. Therefore, such as shown in precedent paragraph the increasing of the applied magnetic field induces a competition between both normal current and intragranular flux motion contributions to the transversal component at high current limit. We can understand the effect of the temperature in a similar manner. With increasing temperature the negative Hall voltage contribution described by motion of intragranular vortices becomes comparable with the positive Hall voltage described by normal state carriers. If intragranular contribution is increased a negative sign to the Hall component with increasing temperature must appear. In the Figure 6 is presented a good experimental evidence for such an interpretation. The $R_{XY}(T)$ curves measured at 70 mA under three different applied magnetic fields show a reduction of the temperature which $R_{XY}$ crosses from positive to negative sign increasing the applied magnetic field.

Finally, in order to compare our results, we discuss the transport properties reported by Khosroabadi *et al.* [28] for a $Gd_{0.90}Pr_{0.10}Ba_2Cu_3O_{7-\delta}$ polycrystalline sample. The electrical resistance as a function of the temperature of that sample presents double transition which can be understood as a granular **character of the** samples [35-38]. Thus, we suggest that the observable positive Hall coefficient in the polycrystalline superconducting samples can be related to granularity properties of weakly coupled superconductors.

## 4  CONCLUSION

It is reported longitudinal and transverse resistance as a function of the temperature measured at low magnetic field for the $Y_{0.55}Pr_{0.45}Ba_2Cu_3O_{7-\delta}$ polycrystalline sample. Hall resistance measurements presented double sign reversal which seems to be related to weak coupling effects. Clockwise hysteresis loops observed in the longitudinal magneto-resistance confirm the hypothesis that $R_{XX}$ and $R_{XY}$ components were measured at low temperature in the regime dominated by the dissipation due to weak coupling between superconducting clusters. It is observed nonzero transverse voltage at zero external magnetic field below superconducting transition ($T_{Ci}$). We also observed that there is a correlation between Hall and longitudinal resistances. The branching point in the $R_{XX}$ component was observed to occur essentially at same temperature in which Hall voltage crosses from negative to positive sign at superconducting state with decreasing temperature. This behavior is interesting and was also observed for other superconductors systems [34]. We suggest that the positive Hall voltage at superconducting state can be related to the quasiparticle currents which agree with our recent discussion about the two-fluid currents flowing in superconducting samples [43].

## ACKNOWLEDGEMENTS

This work has been supported by FAPESP (00/03610-4 and 97/11113-6) and CNPq. The authors would like to thank C. Y. Shigue for his suggestions.

**FIGURE CAPTIONS**

**FIG. 1** – (a) Electrical resistance as a function of the temperature R(T) measured at zero applied magnetic field and different applied currents for the sample with $Y_{0.55}Pr_{0.45}Ba_2Cu_3O_{7-\delta}$ composition. (b) R(T) measured at different applied magnetic fields for the same sample In inset is shown a schematic view of the electrical contacts used during the measurements of the longitudinal ($V_{XX}$) and transversal ($V_{XY}$) voltages.

**FIG. 2** – Magneto-resistance hysteresis loops measured at 4.2 K for three different applied currents.

**FIG. 3** – (a) Longitudinal ($R_{XX}$) and Hall ($R_{XY}$) resistances measured at zero applied magnetic field. (b) $R_{XX}$ and $R_{XY}$ resistance measured at applied magnetic field of $H_A = 4.8$ Oe.

**FIG. 4** – Longitudinal ($R_{XX}$) and Hall ($R_{XY}$) measured under at applied magnetic $H_A = 4.8$ Oe at different applied currents.

**FIG. 5** – Current-voltage characteristic curves of both $R_{XX}$ and $R_{XY}$ components measured at 4.2 K for different applied magnetic fields. Inset highlights the longitudinal component at high applied current limit.

**FIG. 6** – Decreasing of the temperature in which occurs the crossover from positive to negative in the Hall resistance with increasing applied magnetic field.

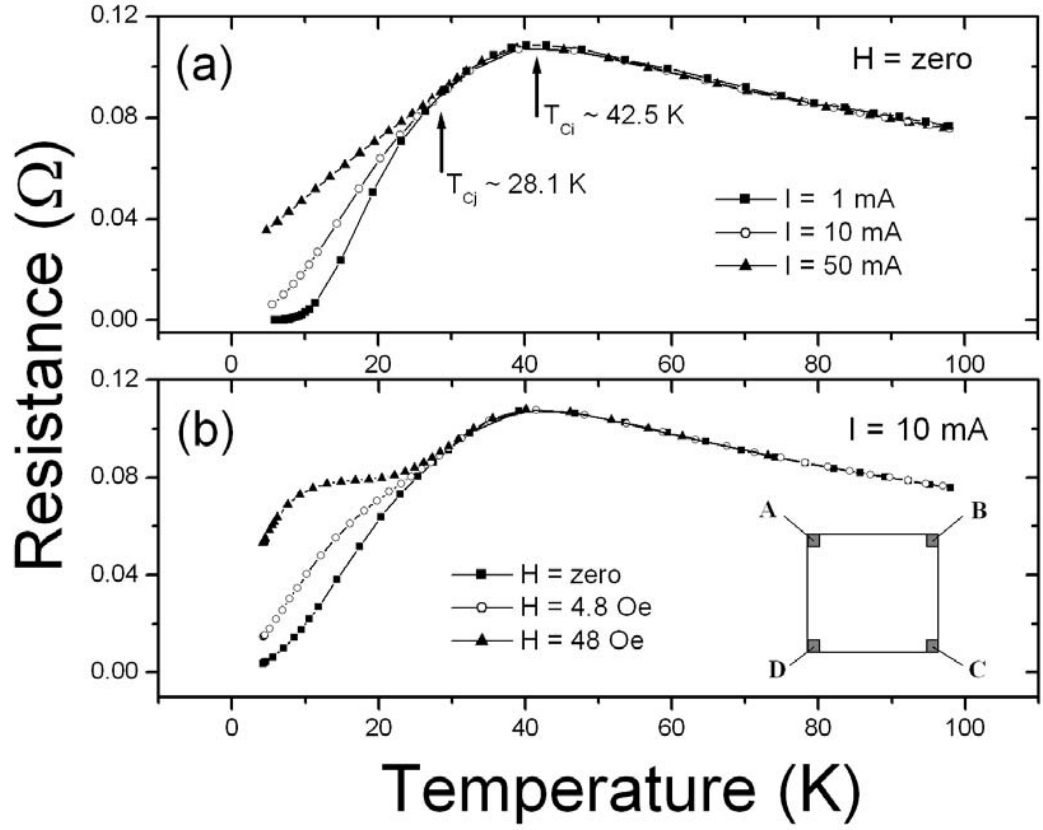

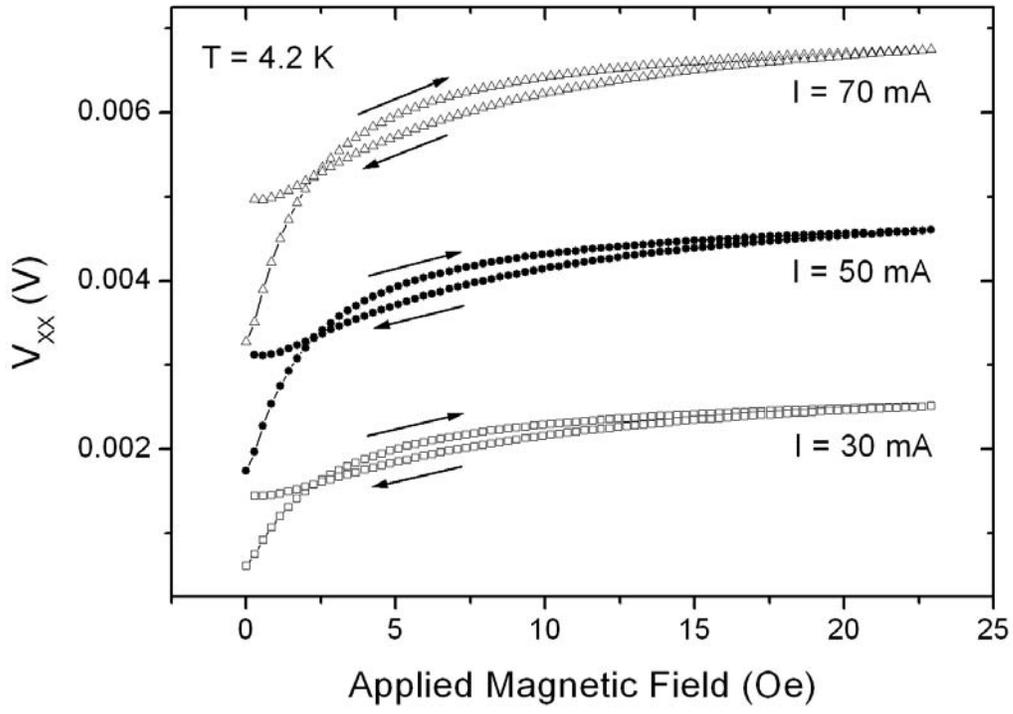

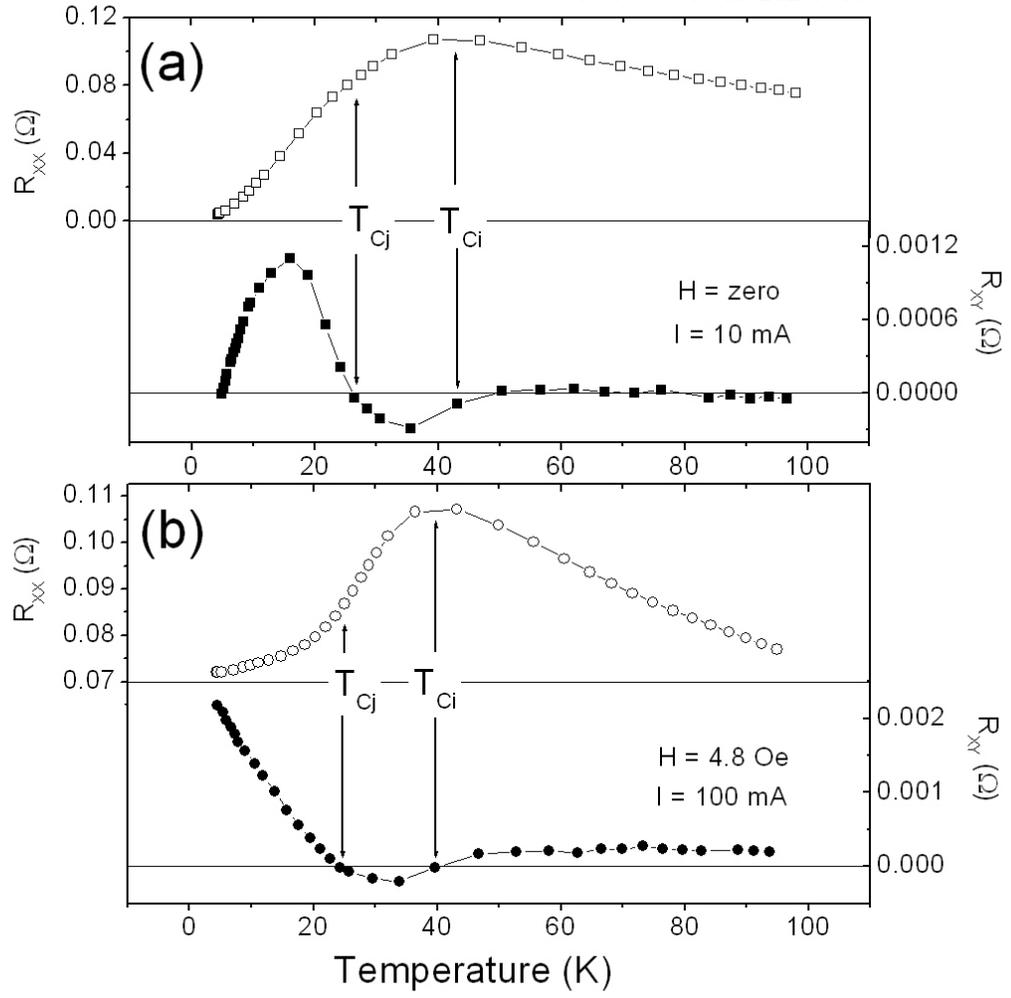

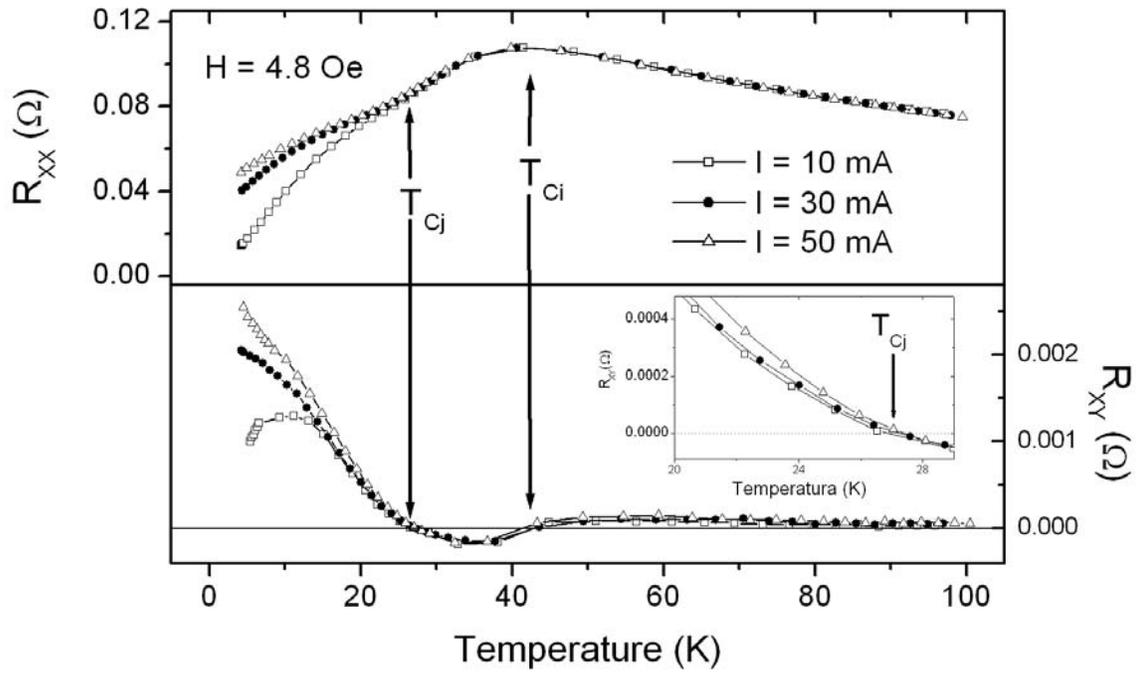

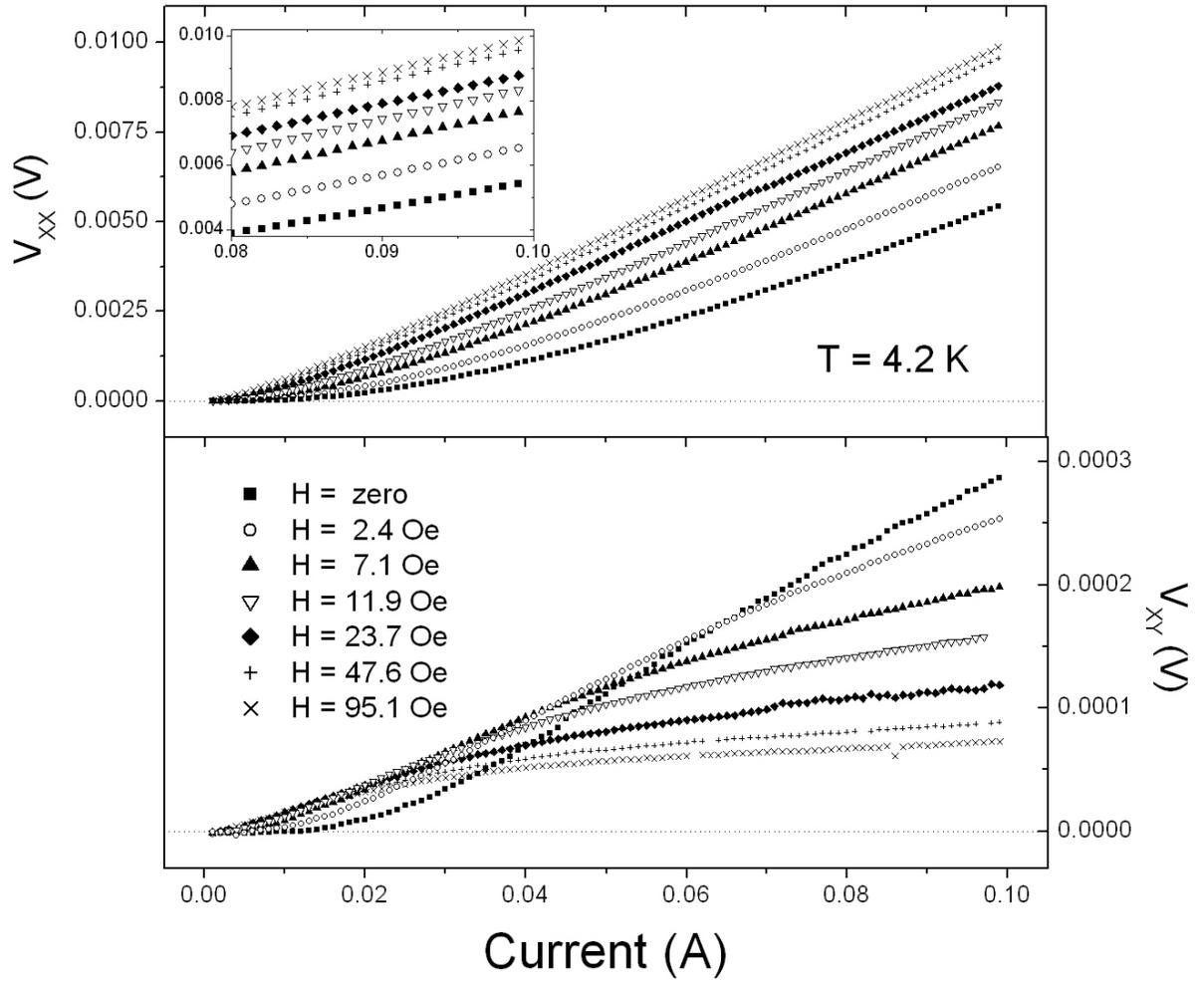

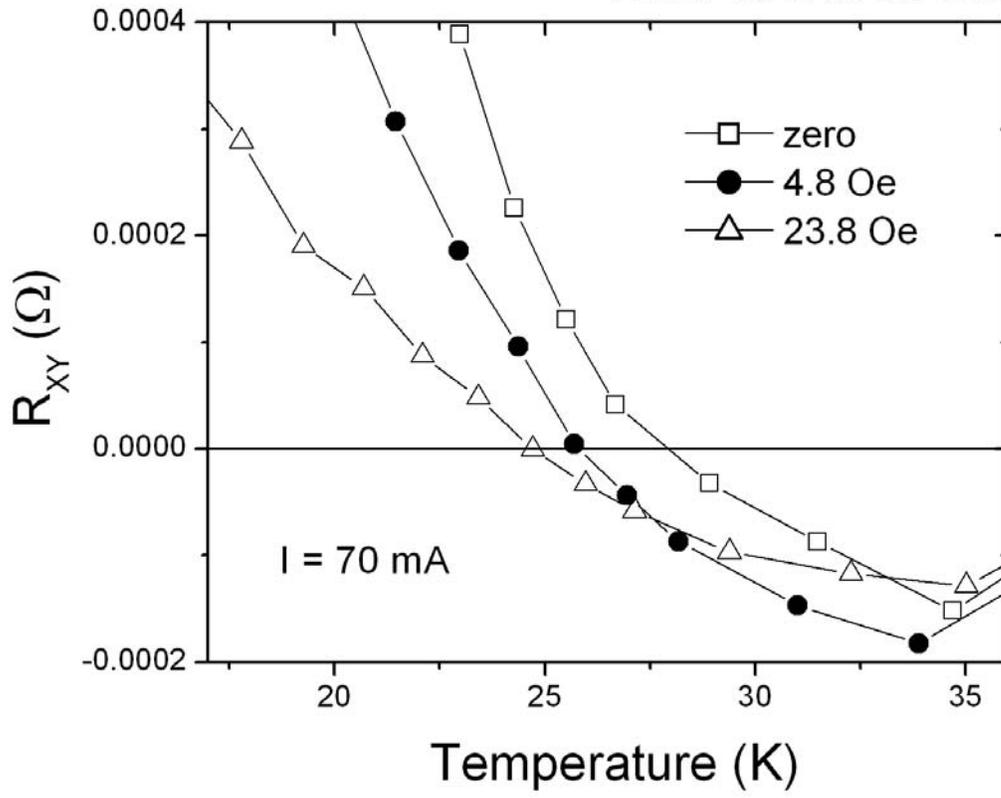

Legend:
□ zero
● 4.8 Oe
△ 23.8 Oe

I = 70 mA